\documentstyle[eqsecnum,aps,epsbox]{revtex}
\renewcommand{\prd}{{\it Phys. Rev. D~}}

\begin{document}
\draft
\title{Cosmological Evolution of Black Holes in Brans-Dicke Gravity}
\author{Nobuyuki Sakai\thanks{%
Electronic address: sakai@ke-sci.kj.yamagata-u.ac.jp}}
\address{Faculty of Education, Yamagata University, Yamagata 990-8560,
Japan}
\author{John D. Barrow\thanks{%
Electronic address: J.D.Barrow@damtp.cam.ac.uk}}
\address{DAMTP, Centre for Mathematical Science, University of\\
Cambridge, Wilberforce Road, Cambridge CB3 0WA, U.K.}
\date{6 February 2001}
\maketitle

\begin{abstract}
We consider a modified ``Swiss cheese'' model in Brans-Dicke theory, and use
it to discuss the evolution of black holes in an expanding universe. We
define the black hole radius by the Misner-Sharp mass and find their exact
time evolutions for dust and vacuum universes of all curvatures.
\end{abstract}

\vskip 5mm
\begin{center}
PACS numbers: 04.70.Bw, 04.50.+h
\end{center}


\section{Introduction}

The evolution of primordial black holes in scalar-tensor theories has been
studied in the literature by several authors \cite{Bar,BC,SST,Kan,Jac,SS}.
Some questions about what happens to a black hole in an expanding universe
when $G$ varies in spacetime in these theories were posed by Barrow \cite
{Bar}, who considered two possible scenarios: (a) the effective
gravitational ``constant'', $G(t),$ at the black hole horizon changes along
with its cosmological evolution so that the size of a black hole is
approximated by $R=2G(t)M$. (b) $G$ remains constant at the black hole event
horizon while it evolves on larger scales; a large inhomogeneity in $G$ is
therefore generated. The case (b) was called ``gravitational memory''
because the black hole remembers the value of $G$ at its formation time. In
either case the observational constraints on the abundance of exploding
primordial black holes deduced from the total radiation backgrounds today
would be modified \cite{BC}.

Scheel, Shapiro and Teukolsky \cite{SST} made numerical analyses of dust
collapse in Brans-Dicke (BD) theory, showing that the surface area of the
event horizon decreases with time, contrary to the case in Einstein theory.
Kang \cite{Kan} gave an analytic explanation for the surface-area decrease
in BD theory. Later, Jacobson claimed that there is no ``gravitational
memory'' effect, by analyzing the evolution of a scalar field $\phi (t,r)$
in Schwarzschild background \cite{Jac}. He found a particular solution of
the scalar wave equation which matches smoothly between the black hole and a
special cosmological background and showed that $\phi $ at the event horizon
evolves along with its asymptotic value $\phi (t,\infty )$. However, this
solution requires a particular cosmological variation of $G(t)$ to occur in
the background and may be special. It was also argued that even if the black
hole mass in the Einstein frame is constant then its mass in the Jordan
frame is time-dependent.

In order to investigate the time-dependence of the black-hole mass, Saida
and Soda constructed a ``cell lattice'' universe in BD theory \cite{SS}.
In their model the universe is first tessellated by identical polyhedrons,
which are then replaced by Schwarzschild black holes. It was shown that the
black-hole mass has an adiabatic time dependence, which is qualitatively
different according to the sign of the curvature of the background universe.

As an extension of Saida and Soda's work, we consider a ``Swiss cheese'' (or
Einstein-Straus) model \cite{Ein} in BD theory and discuss the
evolution of the radius and mass of black holes in an expanding isotropic
background universe. The usual ``Swiss cheese'' model refers to a
cosmological model in which spherical regions in the
Friedmann-Robertson-Walker (FRW) universe are replaced by Schwarzschild
spacetimes. Here we construct such a model in BD cosmology.

\section{Background Universe in Brans-Dicke Theory}

BD theory is described by the action,
\begin{equation}
S=\int d^4x\sqrt{-g}\left[ \frac \phi {16\pi }{\cal R}-\frac \omega {16\pi
\phi }(\nabla _\mu \phi )^2+{\cal L}_m\right] ,  \label{action}
\end{equation}
where $\phi $ is BD field, $\omega $ is BD parameter, and ${\cal L}_m$
is the matter Lagrangian. The variations of equation (\ref{action}) with
respect
to $g_{\mu \nu }$ and $\phi $ yield the field equations:
\begin{eqnarray} \label{EinEq}
{\cal R}_{\mu \nu }-\frac 12g_{\mu \nu }{\cal R} &=&{\frac{8\pi }\phi }%
T_{\mu \nu }+{\frac \omega \phi }\left[ \nabla _\mu \phi \nabla _\nu \phi -{%
\frac 12}g_{\mu \nu }(\nabla \phi )^2\right] +\nabla _\mu \nabla _\nu \phi
-g_{\mu \nu }\Box \phi ,  \\ \label{SFEq}
\Box \phi &=&\frac{8\pi }{2\omega +3}{\rm Tr}T.
\end{eqnarray}

As a background universe, we assume the FRW spacetime:
\begin{equation}
ds^2=-dt^2+a(t)^2\left\{ {\frac{dr^2}{1-kr^2}}+r^2(d\theta ^2+\sin ^2\theta
d\varphi ^2)\right\} ,  \label{frw}
\end{equation}
where $k$ $=0,\pm 1$ determines the spatial curvature. As an energy-momentum
tensor, we introduce a dust fluid:
\begin{equation}
T_{\mu \nu }=\rho u_\mu u_\nu ,
\end{equation}
where $\rho $ and $u_\mu $ are the density and the four-velocity of dust,
respectively. The field equations (\ref{EinEq}) and (\ref{SFEq}) reduce to
the following equations for the background universe:
\begin{equation}
H^2+{\frac k{a^2}}={\frac{8\pi \rho}{3\phi}}-H{\frac{\dot\phi}\phi}
+{\frac \omega 6}\Bigl({\frac{\dot \phi }\phi }\Bigr)^2,
\label{Feq1}\end{equation}
\begin{equation}
\ddot \phi +3H\dot \phi =-{\frac{8\pi \rho }{2\omega +3}},  \label{Feq2}
\end{equation}
where an overdot denotes $d/dt$ and $H\equiv \dot a/a$ is the Hubble
parameter.

In the flat space case, with $k=0,$ we know the general analytic solution
for a dust universe\cite{gur}:
\begin{equation}
a(t)=a_0(t-t_{+})^{\lambda _{+}}(t-t_{-})^{\lambda _{-}},~~\phi (t)=\phi
_0(t-t_{+})^{\kappa _{+}}(t-t_{-})^{\kappa _{-}},  \label{dustflat}
\end{equation}
with constants $\lambda _{\pm }$ and $\kappa _{\pm }$ defined by
\begin{equation}
\lambda _{\pm }={\frac{\omega +1\pm \sqrt{1+2\omega /3}}{3\omega +4}},
~~\kappa _{\pm }={\frac{1\pm 3\sqrt{1+2\omega /3}}{3\omega +4}},
\end{equation}
and $a_0$, $\phi _0$ and $t_{\pm }$ are arbitrary constants.

If we take $t_{+}=t_{-}$, the general solution (\ref{dustflat}) reduces to
the special power-law solution, (used for example by Saida and Soda
\cite{SS}):
\begin{equation}
a(t)=a_0(t-t_0)^{{\frac{2\omega +2}{3\omega +4}}},~~\phi (t)
=\phi _0(t-t_0)^{{\frac 2{3\omega +4}}}.  \label{special}
\end{equation}
where $t_0$ may be set to zero. In the limit of $t\rightarrow \infty $, the
general solution (\ref{dustflat}) converges to the special power-law
solution (\ref{special}). If the present cosmic age $t$ is large enough,
observations cannot constrain the relation between $t_{+}$ and $t_{-}$.
Therefore, we keep $t_{+}-t_{-}$ a free parameter.

In vacuum case ($\rho =0$) there are analytic solutions for all $k$. The
vacuum solution for $k=0$ is expressed as \cite{ohan}
\begin{equation}
a(t)=a_0t^{{\frac 1{3(1+\alpha )}}},~~
\phi =\phi _0t^{{\frac \alpha {1+\alpha }}},  \label{vacflat}
\end{equation}
with
\begin{equation}
\alpha ={\frac{1\pm \sqrt{1+2\omega /3}}\omega },
\end{equation}
where we have omitted the arbitrary constant $t_0$ by fixing the origin of
the time coordinate $t$. Introducing the conformal time $\eta =\int {dt/a}$,
the vacuum solutions for $k=\pm 1$ are expressed as
\begin{eqnarray}
k=+1: &&a(\eta )=(\sin \eta )^{{\frac{1-\lambda }2}}(\cos \eta )^{{\frac{
1+\lambda }2}},\ \phi (\eta )=(\tan \eta )^\lambda ,  \label{vacclose} \\
k=-1: &&a(\eta )=(\sinh \eta )^{{\frac{1-\lambda }2}}(\cosh \eta )^{{\frac{
1+\lambda }2}},\ \phi (\eta )=(\tanh \eta )^\lambda,  \label{vacopen}
\end{eqnarray}
with
\begin{equation}
\lambda =\pm {\frac 3{3+2\omega }}.
\end{equation}

\section{Modified Swiss-Cheese Model}

Now we consider a model for a black hole embedded in the FRW universe. We
replace a sphere in the FRW spacetime with a vacuum region which contains a
black hole. Here ``vacuum'' means $T_{\mu \nu }=0$, and does not imply that
${\cal R}_{\mu \nu }=0$ due to the existence of BD field.

Extending Israel's junction conditions for a singular (or regular)
hypersurface \cite{Isr}, Sakai and Maeda have studied bubble
dynamics in the inflationary universe \cite{SM}. It was found
that one can solve the equations of motion for the boundary without knowing
the interior metric if the interior is vacuum, $T_{\mu \nu }=0$, or has only
vacuum energy (a cosmological constant), $T_{\mu \nu }=-\rho g_{\mu \nu }$.
Applying this method to the present model, we can determine the mass and the
radius of a black hole without specifying the interior metric, as we shall
show below.

Let us consider a spherical hypersurface $\Sigma $ which divides a spacetime
into two regions, $V^{+}$ (outside) and $V^{-}$ (inside). We define a unit
space-like vector, $N_\mu ,$ which is orthogonal to $\Sigma $ and points
from $V^{-}$ to $V^{+}$. In order to describe the behaviour of the boundary,
we introduce a Gaussian normal coordinate system, $(n,x^i)=(n,\tau ,\theta
,\varphi )$, where $\tau $ is chosen to be the proper time on the boundary.
Hereafter, we denote by $\Psi ^{\pm }$ the value of any field variable $\Psi
$ defined on $\Sigma $ by taking limits from $V^{\pm }$.

For the matter field, we consider a dust (or vacuum) medium for $V^{+}$ and
vacuum for $V^{-}$:
\begin{equation}
{T}{_{\mu \nu }}^{+}=\rho u_\mu u_\nu ,~~{T}{_{\mu \nu }}^{-}=0.  \label{Tmn}
\end{equation}
Although we assume a smooth boundary at which there is no surface density,
it is not obvious that this matching is possible at all times. Therefore, we
introduce a surface energy-momentum tensor on the boundary surface,
\begin{equation}
S_{ij}\equiv \lim_{\epsilon \rightarrow 0}\int_{-\epsilon }^\epsilon
dn~T_{ij}={\rm diag}(-\sigma ,~\varpi ,~\varpi ),  \label{Sij}
\end{equation}
where $\sigma $ and $\varpi $ denote the surface energy-density and the
surface pressure of $\Sigma $, respectively.

If we introduce the extrinsic curvature tensor of the world hypersurface
$\Sigma $, $K_{ij}\equiv N_{i;j}$, we can write the junction conditions on
$\Sigma $ as \cite{SM}
\begin{equation}
\lbrack K_{ij}]^{\pm }=-{\frac{4\pi }\phi }\biggl( S_{ij}-{\frac \omega
{2\omega +3}}{\rm Tr}S\gamma _{ij}\biggr),  \label{jc1}
\end{equation}
\begin{equation}
-{\ S_i^j\ }_{|j}=[T_i^n]^{\pm },  \label{jc2}
\end{equation}
\begin{equation}
K_{ij}^{+}S_i^j+{\frac{2\pi }\phi }\biggl\{S_j^iS_i^j-{\frac \omega {2\omega
+3}}({\rm Tr}S)^2\biggr\}=[T_n^n]^{\pm },  \label{jc3}
\end{equation}
where we have defined the jump in any quantity $\Psi $ by the bracket $[\Psi
]^{\pm }\equiv \Psi ^{+}-\Psi ^{-}$ and the three-dimensional covariant
derivative by the vertical bar $|$. The junction condition for BD field
is derived from equation (\ref{SFEq}) as
\begin{equation}
\lbrack \phi _{,n}]^{\pm }=-{\frac{24\pi }{3+2\omega }}{\rm Tr}S,~~\phi
^{+}=\phi ^{-},  \label{jc4}
\end{equation}
which implies that $\phi $ is continuous at $\Sigma $ and inhomogeneous in
$V_{-}$.

The extrinsic curvature tensor of $\Sigma $ in the homogeneous region $V^{+}$
is given by \cite{SM}
\begin{eqnarray}  \label{ex1}
K_\tau ^\tau &=&\gamma ^3{\frac{dv}{dt}}+\gamma vH,~~  \label{ex2} \\
K_\theta ^\theta &=&{\frac{\gamma (1+vHR)}R}={\frac \epsilon R}\sqrt{
1+\left( {\frac{dR}{d\tau }}\right) ^2-R^2\left( H^2+{\frac k{a^2}}\right) },
\end{eqnarray}
where
\begin{equation}
R=a(t)r|_\Sigma ,~~v\equiv a{\frac{dr}{dt}}\Big|_\Sigma ,~~\gamma \equiv
{\frac{\partial t}{\partial \tau }}\Big|_\Sigma ={\ \frac 1{\sqrt{1-v^2}}},~~
{\rm and}~~\epsilon \equiv {\rm sign}({K_\theta ^\theta })={\rm sign}
\left({\frac{\partial R}{\partial n}}\right) .  \label{defs}
\end{equation}

From equations (\ref{Tmn}), (\ref{Sij}), (\ref{jc2}), (\ref{jc3}),
(\ref{ex1})-(
\ref{defs}), we obtain the equations of motion:
\begin{equation}
{\frac{dR}{dt}}={\frac{dr}{d\chi }}v+HR,  \label{eom1}
\end{equation}
\begin{equation}
\gamma ^3{\frac{dv}{dt}}=-\gamma \biggl\{ \Bigl(1-2w\Bigr)vH-{\frac{2w}R}
{\frac{dr}{d\chi}}\biggr\}+{\frac{2\pi \sigma }\phi }\biggl\{1+4w+{\frac{
(1-2w)^2}{(2\omega +3)}}\biggr\}-{\frac{\gamma^2v^2\rho}\sigma},
\label{eom2}
\end{equation}
\begin{equation}
{\frac{d\sigma }{dt}}=-{\frac{2\sigma (1+w)}R}{\frac{dR}{dt}}+\gamma v\rho ,
\label{eom3}
\end{equation}
where $w\equiv \varpi /\sigma $.

Once initial values of $R$, $v$, and $\sigma $ are given, the equations of
motion (\ref{eom1})-(\ref{eom3}) determine their evolution. As discussed in
\cite{SM}, initial values should satisfy the angular component of (\ref
{jc1}),
\begin{equation}
\gamma (1+vHR)-\epsilon ^{-}\sqrt{1+\left( {\frac{dR}{d\tau }}\right)^2
-{\frac{R_{MS}}R}}=-{\frac{8\pi \sigma R}\phi }\left( {\frac{\omega +1+w}
{2\omega +3}}\right) ,  \label{eom0}
\end{equation}
where we have chosen $\epsilon ^{-}=+1$. $R_{MS}$ is defined by
\begin{equation}
R_{MS}\equiv R^{-}(1-g^{\mu \nu }R_{,\mu }^{-}R_{,\nu }^{-}),
\end{equation}
where $R^{-}$ is defined as $R^{-}\equiv \sqrt{g_{\theta \theta }}$ at
$\Sigma $ on the $V^{-}$side. Since the Misner-Sharp mass is defined as \cite
{MS}
\begin{equation}
M_{MS}\equiv {\frac{R^{-}}{2G}}(1-g^{\mu \nu }{R}_{,\mu }^{-}{R}_{,\nu
}^{-})=\frac{R_{MS}}{2G},  \label{MS}
\end{equation}
we call $R_{MS}$ the ``Misner-Sharp radius''. Note that $R_{MS}$ is a purely
geometrical quantity and independent of theories of gravitation.

If we considered a spherical bubble in which there is no black hole (or
singularity), we would have to solve the field equations with a regularity
condition at the centre and the boundary condition (\ref{eom0}), as done in
\cite{SM}. However, because we are interested in black
hole solutions, we do not have to take a central regularity condition into
account. Thus, we can use equation (\ref{eom0}) to determine $R_{MS}$.
Of course, our treatment does not guarantee the existence of a black
hole in the center. We suppose a black hole as the most interesting
object, which can be modeled by appropriate choices of matter and BD
field configuration.

At the initial time, we suppose $v=0$ and $\sigma =\varpi=0$, so $R_{MS}$
is given by
\begin{equation}
R_{MS}={R^3}\left( H^2+{\frac k{a^2}}\right) .  \label{Rms}
\end{equation}
Let us now discuss whether $v$ and $\sigma$ remain zero during the ensuing
evolution. Suppose $w=0$, then the only nontrivial term in equation
(\ref{eom2})
is $\gamma^2v^2\rho /\sigma $. If $v$ and $\sigma$ evolved from zero, then
equation (\ref{eom0}) shows $vH\sim \sigma /\phi$, so that $\rho v^2/\sigma
\sim
\rho v/H\phi $. Therefore, equations (\ref{eom2}) and (\ref{eom3}) guarantee
that,
if $v=\sigma =0$ at a certain time, $v=\sigma =0$ at all time.
Interestingly, this result is true only for the dust case, $\varpi /\sigma
=0$; otherwise the term $(2w/R)(dr/d\chi)$ in equation (\ref{eom2}) would
shift
$v$ from zero.

In the case of Schwarzschild spacetime, the Misner-Sharp radius coincides
with the event horizon. Although this is not necessarily true for general
spacetimes, we speculate that the Misner-Sharp radius is a well-defined
measure of the size of a black hole. In the next section, we calculate the
evolution of $R_{MS}$ for black holes in several background cosmological
models.

\section{Evolution of Black Holes}

The evolution of the Misner-Sharp radius for the $k=0$ dust universe is
given by equations (\ref{dustflat}) and (\ref{Rms}),
\begin{equation}  \label{Rdustflat}
R_{MS}=a_0^3r_0^3\left( {\frac{\lambda _{+}}{t-t_{+}}}+{\frac{\lambda _{-}}{
t-t_{-}}}\right)^2(t-t_{+})^{3\lambda _{+}}(t-t_{-})^{3\lambda _{-}},
\end{equation}
where $r_0$ is the comoving radius of the vacuum region. Equation (\ref
{Rdustflat}) shows that the black hole size decreases with time.

If we define the black hole mass by
\begin{equation}
M_{MS}\equiv {\frac{\phi R_{MS}}2},
\end{equation}
then it coincides with the mass defined by Saida and Soda \cite{SS}. For the
$k=0$ dust universe, we obtain
\begin{equation}
M_{MS}={\frac{a_0^3\phi _0}2}\left( {\frac{\lambda _{+}}{t-t_{+}}}+{\frac{
\lambda _{-}}{t-t_{-}}}\right) ^2(t-t_{+})^{3\lambda _{+}+\kappa
_{+}}(t-t_{-})^{3\lambda _{-}+\kappa _{-}}.  \label{Mdustflat}
\end{equation}
It is easy to see that  equation (\ref{Mdustflat}) reduces to $M_{MS}=$
constant,
if we choose $t_{+}=t_{-}$, which is the same result as that found by Saida
and Soda \cite{SS}. They also showed $M_{MS}$ increases for $k=+1$ and
decreases for $k=-1$, and concluded that the evolution of the mass depends
qualitatively on the sign of the curvature of the universe. We should note,
however, that their conclusion is true only for the special case
$t_{+}=t_{-} $, or equivalently, only for the asymptotic behavior of $M_{MS}$
at $t\rightarrow \infty $. Our results give the general solution for all
times.

Next, let us consider the scalar-field dominated (vacuum) universe. The
time-dependent solutions for $R_{MS}$ and $M_{MS}$ in flat, open, and closed
universes are given by
\begin{eqnarray}
k=0: &&R_{MS}={\frac{a_0^3r_0^3}{9(1+\alpha )^2}}t^{{\frac{-1-2\alpha }
{1+\alpha }}}, \\
\ &&M_{MS}={\frac{a_0^3r_0^3\phi _0}{18(1+\alpha )^2}}t^{-1}, \\
k=+1: &&R_{MS}={\frac{r_0^3}2}(\cos \eta )^{-{\frac{3+\lambda }2}}(\sin \eta
)^{-{\frac{1+\lambda }2}}(\cos 2\eta -\sin 2\eta -\lambda ), \\
\ &&M_{MS}={\frac{r_0^3}4}(\tan \eta )^\lambda (\cos \eta )^{-{\frac{
3+\lambda }2}}(\sin \eta )^{-{\frac{1+\lambda }2}}(\cos 2\eta -\sin 2\eta
-\lambda ), \\
k=-1: &&R_{MS}={\frac{r_0^3}2}(\cosh \eta )^{-{\frac{3+\lambda }2}}(\sinh
\eta )^{-{\frac{1+\lambda }2}}(\cosh 2\eta -\sinh 2\eta -\lambda ), \\
\ &&M_{MS}={\frac{r_0^3}4}(\tanh \eta )^\lambda (\cosh \eta )^{-{\frac{
3+\lambda }2}}(\sinh \eta )^{-{\frac{1+\lambda }2}}(\cosh 2\eta -\sinh 2\eta
-\lambda ).
\end{eqnarray}
If we take $\omega >500$, as applies to the universe today, then both
$\alpha \sim O(\omega ^{-\frac 12})$ and $\lambda \sim O(\omega ^{-1})$ will
be negligible. We see that $R_{MS}$ and $M_{MS}$ both decrease with
increasing time, except in the contracting phase of the $k=+1$ universe. In
a more general scalar-tensor theory, with non-constant $\omega(\phi)$ we
would expect similar effects to arise and it would be possible for large
changes in $\omega(\phi)$ to occur in the very early universe despite very
slow evolution towards a very large value of $\omega >1000$ today. However,
such a calculation should be performed for a black hole in radiation and
dust-dominated universes.

\section{Discussion}

We have constructed a modified ``Swiss cheese'' model in Brans-Dicke theory
and discussed the evolution of black holes for dust and vacuum universes. We
defined the size of a black hole $R_{MS}$ by the Misner-Sharp mass, and
found that this size always decreases as long as the universe is in an
expanding phase. Although we have not specified the metric around
the black hole, the black hole mass and radius that we have obtained
coincide with those used by Saida and Soda \cite{SS}, who assumed a
Schwarzschild-like metric. This shows that their ansatz of the
Schwarzschild-like metric does not introduce a specialization of the
problem.

Let us make a discussion about the definition of mass. In general it is not
obvious which definition is the most appropriate among many definitions
of quasi-local mass. However, in a spherically symmetric and
asymptotically flat spacetime, the Misner-Sharp mass coincides with the
ADM mass at spatial infinity, and with the Bondi-Sachs mass at null
infinity, as Hayward proved \cite{Hay}. Although the asymptotic
region in our model is not flat but expanding, if the vacuum region inside
$\Sigma$
is much larger than the black hole, we can regard the regions just
inside $\Sigma$ as quasi-asymtotical-flat and hence the above
coincidence makes sense. Among the three the Misner-Sharp mass is the
simplest to calculate.

We have not investigated the behavior of the interior vacuum
region, which was partially investigated by Scheel et al.\ \cite{SST} and
Jacobson \cite{Jac}. Their
analyses suggest that the scalar field increases at the horizon and hence
the horizon area decreases with time as the background universe expands.
Although Jacobson's particular solution of the scalar wave equation, with
very fast variation of $\phi (t,\infty)\propto t$ in the background
universe, look unrealistic, we speculate that his conclusion may be
unchanged for a more
realistic background model because homogenisation of BD field
occurs more effectively for slower time-variation of $\phi (t,\infty)$ \cite
{Sak}.
These studies, however,  indicate that there are still uncertainties to be
resolved as to the behaviour of black holes in background universes with
arbitrary time variations in $G$. A detailed numerical study of a spherical
collapse in a background Friedmann universe will answer some of these
outstanding questions.

\vskip 5mm \noindent
{\large {\bf Acknowledgments}}\\

N.\ S.\ thanks T. Chiba, T. Harada, J. Koga, K. Maeda, H. Saida, H. Shinkai
and T. Tanaka for useful comments, and also DAMTP, University of Cambridge,
where a part of this work was carried out. This work was partially supported
by JSPS Programs for Research Abroad 1999 and the Grant-in-Aid for
Scientific Research Fund of Monbu-Kagakushou (No.\ 13740139).

\end{document}